\documentclass[10pt,a4paper]{article}

\usepackage{graphicx}
\usepackage{latexsym}
\usepackage{mathrsfs}
\usepackage{euscript}
\usepackage[centertags]{amsmath}
\usepackage{amsfonts}
\usepackage{amssymb}
\usepackage{amsthm}
\usepackage{newlfont}
\usepackage{anysize}
\usepackage{setspace}

\title{Cryptanalysis of A Secure Remote User Authentication
Scheme Using Smart Cards}
\author{Tanmoy Maitra\\ E-mail: tanmoy.maitra@live.com\\Department of Computer Science and Engineering\\Jadavpur University\\Jadavpur 700032, India}
\date{}
\begin{document}
 \maketitle
\abstract{Smart card based authentication schemes are used in
various fields like e-banking, e-commerce, wireless sensor
networks, medical system and so on to authenticate the both remote
user and the application server during the communication via
internet. Recently, Karuppiah and Saravanan proposed an
authentication scheme which is based on password and one-way
cryptographic hash function. They have used a secure identity
mechanism i.e., users' and server's identity are not public. Thus,
the user and the server do not send their identity directly to
each other during communications. In this paper, we have found out
that their scheme does not overcome the reply attack and also
there is a fault in the login phase, which makes their scheme is
not perfect for practical use.} 

\section{Introduction}
Smart card based mutual authentication system provides a facility
where both communicators can verify each other during the online
services. For this purpose, in single server environment based
authentication system, the users do their registration for one
time to a server to get services from that server for several
times. After registration, each user gets his/her smart card form
the server. By using their smart card, users get services from the
server via public channel through internet. A good user
authentication scheme should follow the following properties:
\begin{itemize}
\item Efficient login phase so that, the smart card can recognize
the wrong inputs from the users before going to send login message
to the server.
    \item Users can freely change their password with or without
    help from the server.
    \item The strong mutual authentication should satisfy.
    \item Perfect forward secrecy should hold so that, the computed shared session
    key is only known to the user and the server during that
    communication session.
    \item Communication overhead must be less so that, the
    authentication scheme provides good efficiency.
\item The design scheme should resist the all possible attacks
such as, insider attack, guessing attack, smart card stolen
attack, forgery attack, man-in-middle attack and so on.
\end{itemize}     There are many password based
authentication systems~\cite{1,2,3,4} in the literature. In 2012,
Chen et al. \cite{chen} proposed a robust smart card-based remote
user password authentication scheme. In 2013, Kumari and Khan
\cite{knk} showed that Chen et al.'s scheme cannot resist
impersonation attacks and insider attacks, and they then presented
an improved scheme. In the same year, Li et al. \cite{li} also
showed that Chen et al.'s scheme cannot ensure perfect forward
secrecy and that it cannot detect incorrect passwords in the login
phase, and they then proposed an improved scheme. Recently,
Karuppiah and Saravanan~\cite{kns} proposed a password based user
authentication scheme in single server environment to provide the
robustness of the authentication system. They claim that their
scheme follows the above properties which make their scheme better
than related schemes. But, in this paper, we have shown that there
is a fatal error in login phase of their scheme so that, their
scheme is no more applicable for practical use. Besides, we have
pointed out the disadvantage in login phase which may mount replay
attack on their scheme. \par The rest of the paper is organized as
follows: Section~\ref{scheme} presents the brief review of
Karuppiah and Saravanan's Scheme. Section \ref{ana} shows the
weaknesses of Karuppiah and Saravanan's Scheme. Finally, the
conclusion appears in Section~\ref{con}.

\section{Review of Karuppiah and Saravanan's Scheme}\label{scheme}
In this section, we will briefly discuss the Karuppiah and
Saravanan's scheme~\cite{kns}, in which we try to use the same
notations as presented in their paper. Their scheme consists of
five phases namely, initialization phase, registration phase,
login phase, authentication phase and password change phase.
\subsection{Initialization Phase}
A server $S$ selects two large prime numbers $p$ and $q$. Further,
the server chooses a generator $g$ of a finite field in $Z^*_p$.
Then, the server computes $n$ =  $p\times q$ and $\phi(n)$ =
$(p-1)\times (q-1)$. Then, the server chooses an integer number
$e$ such that $gcd (e, \phi(n))$ = 1 and $1 < e < \phi(n)$. The
server computes an integer $d$ such that $d$ = $e^{-1} ~mod~
\phi(n)$ and $y$ = $g^d ~mod~ n$. Finally, the server declares $y$
as a public key of it and keeps $<d,p,q>$ as secret.
\subsection{Registration Phase}
When a new user $U_i$ wants to register to access the server $S$,
this phase is invoked. The user $U_i$ freely selects his/her
identity $ID_i,$ password $PWD_i$ and a random number $b$. Then,
the $U_i$ computes $h(b\oplus PWD_i)$ and sends $\langle
ID_i,h(b\oplus PWD_i)\rangle$ to the server $S$ for registration.
After receiving the registration message $\langle ID_i,h(b\oplus
PWD_i)\rangle$, the server verifies credential of identity $ID_i$.
If it finds $ID_i$ in its database, that means, $ID_i$ is
registered with some other user, and the server asks for a new
identity to the user $U_i$. Otherwise, the server $S$ issues a
smart card that contains public parameters $\langle C_{in}, B_1,
g, y, n, h(\cdot) \rangle $  for the user $U_i$ after computing
$B_1$ = $h(ID_i)^{h(b\oplus PWD_i)} ~mod~ n$ and $C_{in}$ =
$y^{h(d\parallel T_R\parallel ID_i)+h(b\oplus PWD_i)}~mod ~n$,
where $d$ and $T_R$ are the server's secret key and the
registration time and date of user $U_i$ respectively. Further,
the server creates an entry for $U_i$ in the database and stores
an encrypted form of $(ID_i, T_R$) in this entry. Finally, the $S$
sends the smart card to the user $U_i$. After getting the smart
card, the user inserts the random number $b$ into the memory of
the smart card.
\subsection{Login Phase}
In this phase, the user inserts his/her smart card to the terminal and provides his/her identity $ID_i^*$ and password $PWD_i^*$ to the terminal.
The terminal or smart card computes the following steps:
\begin{enumerate}
    \item The smart card computes $B_1^*$ =
$h(ID_i^*)^{h(b\oplus PWD_i^*)} ~mod~ n$ and compares $B_1== B_1^*$. If it holds good, the smart card computes the following steps; otherwise rejects the user $U_i$.

    \item The smart card computes $B_2$ = $g^j~mod~n$, $B_3$ =
    $y^j~mod~n$,  $C$ = $ID_i\oplus h(B_2\oplus B_3)$,
    $C_{in}^{\prime}$ = $C_{in}\times y^{-h(b\oplus
    PWD_i^*)}~mod~n$ (= $y^{h(d\parallel T_R\parallel
    ID_i)}~mod~n$) and $M$ = $h(C_{in}^{\prime}\parallel C)$, where a random number $j$ is generated by the smart card.
    Then, the smart card sends a login request message $\langle B_2,M,C\rangle
    $ to the server $S$.
    \item After receiving the login request message $\langle B_2,M,C\rangle
    $ from the user $U_i$, the server $S$ computes $B_3^{\prime}$
    = $(B_2)^d~mod~n$ (= $y^j~mod~n$), derives $ID_i$ = $C\oplus h(B_2\oplus
    B_3^{\prime})$ and checks the validity of the user $U_i$. If
    it is valid proceeds to the next steps; otherwise rejects the
    login message.
    \item The server $S$ computes $C^*$ =  $y^{h(d\parallel T_R\parallel
    ID_i)}~mod~n$, $M^*$ = $h(C^*\parallel C)$ and checks
    $M^*==M$. If the equality holds, proceeds to next steps;
    otherwise rejects the login message.
    \item The server $S$ computes $t$ = $h(T_s\oplus ID_i\oplus ID_s\oplus
    B_3^{\prime})$, $C_1$ = $(C^*)^{r+t}~mod~n$, where $T_s$ and
    $r$ are the current time and date of the server $S$ and a
    random number generated by the server $S$. Then, the server sends
    a reply message $X$ = $\langle h(C_1), r,T_s \rangle $ to the user
    $U_i$ at time $T_s$.
    \item After receiving the reply message $X$ = $\langle h(C_1), r,T_s
    \rangle$ from the server $S$ at time $T$, the smart card checks whether $(T-T_s)\le \triangle T$ or not.
    If it holds good, the smart card proceeds to next; otherwise rejects the reply
    message of the server $S$.
    \item The smart card computes $t^*$ = $h(T_s\oplus ID_i\oplus ID_s\oplus
    B_3)$, $C_2$ = $(C_{in}^{\prime})^{r+t^*}~mod~n$ and checks
    $h(C_2)==h(C_1)$. If it holds good, the smart card proceeds to
    next; otherwise rejects the reply message of the server $S$.
    \item The smart
card computes $M_1$ = $(h(C_2\oplus ID_i))^T~mod~ n$, where $T$ is
the current time and date of the smart card reader clock. The
smart card sends a message $Z$ = $\langle M_1, T\rangle $ to the
server $S$. \end{enumerate} \subsection{Authentication Phase}
After receiving the message $Z$ = $\langle M_1, T\rangle$ from the
user $U_i$ at time $T_s$, the server checks whether $(T_s-T)\le
\triangle T$ or not.
    If it holds good, the server performs the following steps; otherwise rejects the
    the
    message $Z$ = $\langle M_1, T\rangle$ of the user $U_i$.
    \begin{enumerate}
        \item The server computes $M_2$ = $(h(C_1\oplus ID_i))^T~mod~
        n$ and checks $M_1==M_2$. If it is true, the server
accepts the login request and grants permission to the user $U_i$;
otherwise, the server rejects the login request.
        \item After successful mutual authentication, the user $U_i$ and the server
        $S$
independently compute the common session key as $S^U_{Key}$ =
$h(ID_i \parallel ID_S \parallel C_2)$ and $S^S_{Key}$ = $h(ID_i
\parallel ID_S\parallel C_1)$ respectively.

    \end{enumerate}

    \section{Cryptanalysis of Karuppiah and Saravanan's
    Scheme}\label{ana}
 In this section, we will analyze the Karuppiah and Saravanan's scheme \cite{kns} and will demonstrate the
disadvantage and the faulty login phase.
\subsection{Faulty Login Phase}
In the Karuppiah and Saravanan's scheme, identity $ID_i$ of the
user $U_i$ and also the identity $ID_s$ of the server $S$ are not
public that means, user $U_i$'s identity $ID_i$ is not stored into
his/her smart card directly and also the user $U_i$ does not send
his/her identity $ID_i$ directly with the login message to the
server $S$ in login phase. For this purpose, to verify the
legitimate user $U_i$, the server $S$ stores an encrypted form of
$(ID_i, T_R)$ in its database during the registration phase and
when a login message is received by the server, it computes
$B_3^{\prime}$
    = $(B_2)^d~mod~n$ (= $y^j~mod~n$), derives $ID_i$ = $C\oplus h(B_2\oplus
    B_3^{\prime})$ and checks whether the derived $ID_i$ is present into its database or
    not. If the derived $ID_i$ is found into its database, the
    server computes the remaining steps of the login phase;
    otherwise, rejects the user $U_i$. The above procedure shows
    that unless the identity $ID_i$ of the user $U_i$ is derived,
    the server can not recognize the user $U_i$. Similarly, to
    recognize the server $S$ with its identity $ID_s$, the user
    must know the identity $ID_s$ of the server. But, there is no
    procedure to know server's identity for the user $U_i$ because,
    the $ID_s$ is not public and also the server $S$ does not send
    $ID_s$ with the reply message directly to the user $U_i$ in
    the login phase. The server sends reply message $\langle h(C_1), r,T_s
    \rangle$ by computing $C_1$ = $(C^*)^{r+t}~mod~n$, where $t$ = $h(T_s\oplus ID_i\oplus ID_s\oplus
    B_3^{\prime})$, $r$ is a random number chosen by the server
    and $T_s$ is the current time and date of the server $S$.
    According to the Karuppiah and Saravanan's scheme, after receiving the reply message $\langle h(C_1), r,T_s
    \rangle$ from the server $S$, the user $U_i$ computes $t^*$ = $h(T_s\oplus ID_i\oplus ID_s\oplus
    B_3)$, where $T_s$ is known to the user from the reply
    message, $U_i$ knows his/her identity $ID_i$, $B_3$ (= $y^j~mod~n$) is also
    known to the user because, he/she computes this parameter
    during the login phase and $ID_s$ is unknown to the user
    $U_i$.  Though the user $U_i$ does not know $ID_s$, he/she computes $t^*$ = $h(T_s\oplus ID_i\oplus ID_s\oplus
    B_3)$. This is a fatal error of the Karuppiah and Saravanan's
    scheme. Thus, the $U_i$ can not compute $t^*$ = $h(T_s\oplus ID_i\oplus ID_s\oplus
    B_3)$ without knowing $ID_s$. Hence, the Karuppiah and Saravanan's
    scheme is not perfect for practical use.
\subsection{Disadvantage}
The login request message $\langle B_2,M,C\rangle $ is depended on
only a random number $j$ generated by the smart card as $B_2$ =
$g^j~mod~n$, $C$ = $ID_i\oplus h(B_2\oplus B_3)$ (= $ID_i\oplus
h(g^j~mod~n\oplus y^j~mod~n)$, as $B_3$ = $y^j~mod~n$) and $M$ =
$h(C_{in}^{\prime}\parallel C)$ (= $h(y^{h(d\parallel T_R\parallel
ID_i)}~mod~n \parallel C$)), where $T_R$ is the registration time
and date of the user $U_i$. $T_R$ is a fixed parameter because,
one user can register to the server only one time with his/her
identity $ID_i$. But, the user can access the server for several
times after performing the valid registration procedure only one
time. We assume that the previous login request message of the
previous session between a user $U_i$ and the server $S$ is stored
in the server end. After getting login request message form $U_i$
for a new session, $S$ checks the current login request message
with previous login request message. If they are same, $S$ rejects
the current login request message to avoid replay attack. An
adversary traps the login request messages for some sessions
$ST_1, ST_2,\dots ,ST_m$ with $ST_1<ST_2<\cdots<ST_m$, where
$ST_i<ST_j$ means $ST_i$ is a pervious session than $ST_j$.
Suppose, the adversary sends the trapped login message to $S$ in
any session from $\{ST_1, ST_2,\dots ,$ ${ST}_{m-1}\}$ to the next
session, say, ${ST}_{m+1}$. $S$ accepts the login request message
of the adversary. To resist replay attack in Karuppiah and
Saravanan's scheme, the server has to store all the previous login
request messages for all the users to check with the current login
request message. It is not an efficient technique where server
takes more time to search and compare the messages only to resist
reply attack.
\section{Conclusion and Future Scope}\label{con}
We have shown that Karuppiah and Saravanan's scheme has a fatal
error in login phase so that their scheme is impractical for real
world application. Further, we have also shown the disadvantage of
their scheme. In future, we will improve their scheme to overcome
the fatal error in login phase as well as eliminate the
disadvantage of their scheme.

 \end{document}